# Performance Evaluation of Low Power MIPS Crypto Processor based on Cryptography Algorithms

## Kirat Pal Singh[1], Dilip Kumar[2]

Academic & Consultancy Services Division
Centre for Development of Advanced Computing (C-DAC)
A scientific society of the ministry of Communications & Information Technology (Govt. Of India)
Mohali-160071, Punjab, India

**ABSTRACT**
This paper presents the design and implementation of low power 32-bit encrypted and decrypted MIPS processor for Data Encryption Standard (DES), Triple DES, Advanced Encryption Standard (AES) based on MIPS pipeline architecture. The organization of pipeline stages has been done in such a way that pipeline can be clocked at high frequency. Encryption and Decryption blocks of three standard cryptography algorithms on MIPS processor and dependency among themselves are explained in detail with the help of a block diagram. Clock gating technique is used to reduce the power consumption in MIPS crypto processor. This approach results in processor that meets power consumption and performance specification for security applications. Proposed Implementation approach concludes higher system performance while reducing operating power consumption. Testing results shows that the MIPS crypto processor operates successfully at a working frequency of 218MHz and a bandwidth of 664Mbits/s.

## 1. Introduction

Today's digital world, Cryptography is the art and science that deals with the principles and methods for keeping message secure. Encryption is emerging as a disintegrable part of all communication networks and information processing systems, involving transmission of data. Encryption is the transformation of plain data (known as plaintext) into intengible data (known as cipher text) through an algorithm referred to as cipher. There are two classes of Key Based Encryption Algorithm: Symmetric and Asymmetric algorithms. The most commonly used technique for producing confidentiality in data transmission is symmetric algorithm. This algorithm performs various mathematical and logical functions on the plaintext using the same key where as asymmetric algorithm use different keys for encryption and decryption process. In both algorithms, the key is essential part of encryption and decryption process which provides secure data traffic among Sender and Receiver.

The MIPS is simply known as Millions of instructions per second and is one of the best RISC (Reduced Instruction Set Computer) processor ever designed. MIPS architecture is employed in a wide range of applications. The architecture remains the same for all MIPS based processors while the implementations may differ [1]. There is a 16- bit RSA cryptography MIPS cryptosystem have been previously designed [2]. Some adjustments and minor improvements in the MIPS pipelined architecture design are made using authenticating devices [3] such as Data Encryption Standard [DES], Triple-DES and Advanced Encryption Standard [AES] to protect data transmission over insecure medium. High speed MIPS processor possesses Pipeline architecture to speed up the processing as well as increase the frequency and performance. A MIPS based RISC processor was described in [4]. It consists of basic five stages of pipelining that are Instruction Fetch, Instruction Decode, Instruction Execution, Memory Access and Write Back. These five pipeline stages generate 5 clock cycles processing delay and several Hazards during the operation [2]. These pipelining Hazard are eliminates by inserting NOP (No Operation Performed) instruction which generate some delays for the proper execution of instruction [4]. The pipelining Hazards are of three types: data, structural and control hazard. These hazards are handled in the MIPS processor by the implementation of Forwarding Unit, Pre-fetching or Hazard detection unit, Branch and Jump Prediction Unit [2]. The Forwarding unit is used for preventing data hazards which detects the dependencies and forward the required data from the running instruction to the dependent instructions [5]. Stall occurs in the pipelined architecture when the consecutive instruction uses the same operand as that of the instruction and requires more clock cycles for execution. This reduces the performance. To overcome this situation, Instruction Pre-fetching Unit is used which reduces the Stalls and improves performance. The control hazard occurs when a branch prediction is mistaken or in general, when the system has no mechanism for handling control hazards [5]. The control hazard is handled by two mechanisms: Flush mechanism and Delayed jump mechanism. The branch and jump prediction unit uses these two mechanisms for preventing control hazards. The flush mechanism runs instruction after a branch and flushes the pipe after the misprediction [5]. Frequent flushing may increase the clock cycles and reduce performance. In the delayed jump mechanism, Specific numbers of NOP's are pipelined after the Jump instruction to handle the control hazard. The branch and jump prediction unit placement in the pipelining architecture may affect the critical or the longest path. The standard method of increasing performance of the processor is to detect the longest path and design hardware that results in minimum clock period.

This paper is organized as follows. MIPS crypto system is explained in Section II. A mathematical background for the cryptography algorithms are explained in sub-sections and can be found in [3] [7] [9] [12-15]. The system architecture specifications and implementation methodology are explained in Section III. Hardware implementation design and Instruction set of MIPS including new instructions in detail with corresponding diagrams are shown in sub-section. The implementation results of pipeline stages are shown in Section IV. Design performance and area requirement of MIPS crypto pipeline processor and their Verification & synthesis report are described in sub sections. The discussion and future work are described in Section V. The conclusion is provided in Section VI.

## 2. MIPS Crypto Processor Architecture

The single chip MIPS crypto processor (shown in Fig. 1) consists of various components like Datapath, Data I/O unit, Control Unit, Memory unit, Crypto Specific Unit, Dependency Resolver and Arithmetic Logic Unit. The dedicated data processing block consist of Datapath and Crypto IP core (coprocessor) that performs the 128-bit AES cipher operation and a 64-bit DES/TDES cipher or decipher operation. Advanced Encryption Standard (AES) algorithm operates on 128bits block size by using cipher keys with lengths 128, 192 and 256 bits for encryption process respectively. The incoming data and key are stored in a matrix called state matrix and all the operations are performed over the state matrix [6]. Data Encryption Standard (DES) and Triple DES is a Symmetric crypto algorithm, which operates on 64-bit block size with 16 rounds. The input plaintext, cipher keys and output cipher text are of 64-bit. The main operation in DES and TDES is bit permutation and substitution in one round which is performed by the permutation unit. Datapath processing unit performs the 5 stages pipelining process inside the processor. It consists of Program Counter, 32-bit General Purpose Registers, Key Register and Sign Extender Unit. The program counter unit updates the values available at its input bus at every positive edge clock cycle and also fetches the next instruction from the instruction ROM memory. The registers are read from the General purpose register and the opcode is passed to the control unit which asserts the required control signals. Sign extension is used for calculating the effective address. The data and instruction memory have capability of storing 256 bytes and each byte is referred by the address in between 0 to 256. The address is represented by 8-bits.

The MIPS controller is the main core of the architecture which consists of control unit and ALU control signal unit. The function of controller is to controls the dedicated crypto block and performs the interface and specific operation with the external devices such as Memory, I/O bus interface controller. Single control unit controls the activities of other modules according to the instruction stored inside memory. The crypto specific block executes various other private and public key algorithms such as RSA, DSA, elliptic curve and IDEA with other application programs such as user authentication programs.

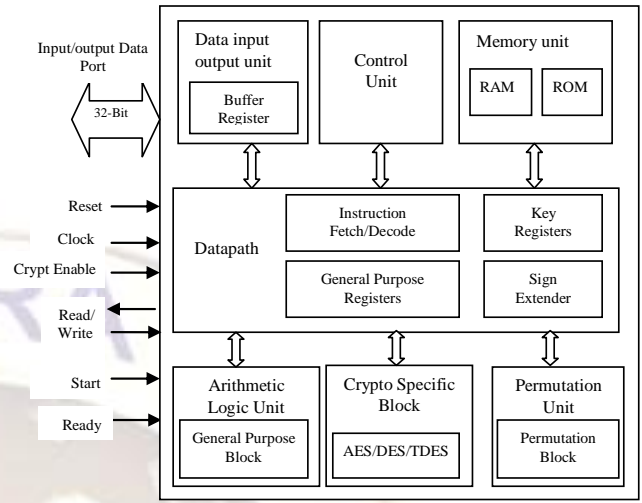

**Fig. 1.** MIPS crypto processor architecture

The arithmetic logic unit (ALU) performs the NOP (no operation), addition, subtraction, OR, NOR, set less than, shift left logic operation. The data and address calculations for load and store instruction are performed by ALU. The Load and Store instructions write to and read from the RAM memory in the memory unit while the ALU results and the data read from RAM are written in to the register file by the register type and Load instruction respectively. Data I/O has two different external interfaces which stored data initially at buffer registers or move data to output. The bit permutation operation has a big process part in DES and TDES algorithms as it improves diffusion properties. The incoming data is subjected to some bit position according to the permutation type. The dependency resolver block has a function to avoid stall by rearranging the instruction sequence and checking the successive instruction for their stall possibility by comparing their operands. This module handles both stalling as well as data forwarding of previous stage. In case of data dependency between two consecutive instructions the receiving instruction waits for one clock cycle. Thus dependency resolver controls the data forwarding in pipeline stages.

### 2.1 Data Encryption Standard

Data Encryption Standard (DES) algorithm uses the complicated logical function such as non-linear permutation and substitution. In this algorithm, there are 16 rounds of identical operation and in each round, 48-bit sub keys are generated, and substitution using S-box, bitwise shift, and XOR (exclusive –OR) operation are performed. The algorithm is designed to encrypt and decrypt blocks of data consisting of 64-bit using 56-bit key. Sometimes the key is considered as 64-bits in length for computational purpose (but only 56bits are used for conversion purpose and rest bits are used for parity checking). DES acts on 64-bit block of the plaintext, involving 16 rounds of permutations, swap, and substitutes as shown in Fig. 2. The standard includes labels describing all of the selection, permutation and expansion operations mentioned below; these aspects of the algorithm are not secrets. The basic DES steps are:

(1) The 64-bit block to be encrypted undergoes an initial permutation (IP), where each bit is moved to a new bit position; e.g., the 1st, 2nd and 3rd bits are moved to the 58th, 50th and 42nd position, respectively.

(2) The 64-bit permuted input is divided into two 32-bit blocks, called left and right, respectively. The initial values of the left and right blocks are denoted L0 and R0.

(3) These are then 16 rounds of operation on L and R blocking. During each iteration (where n ranges from 1 to 16).

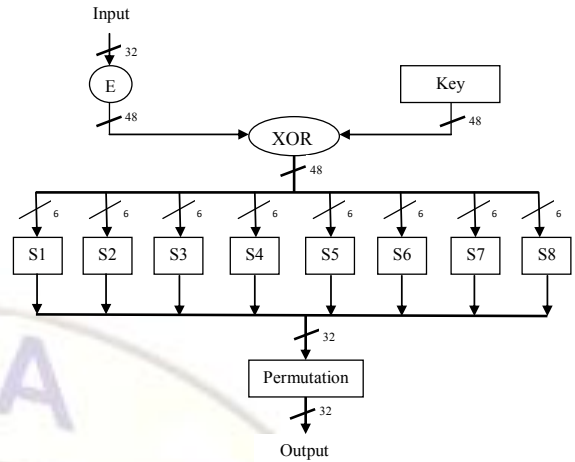

**Fig.3.** DES Round (F[R,K] box) Detail

(4) The results from the final DES round- i.e., L16 and R16 are recombined into a 64-bit value and fed into an inverse initial permutation (IP$^{-1}$). At this step, the bits are rearranged into their original positions, so that 58th, 50th, and 42nd bits, for example, one moved back into the 1st, 2nd and 3rd positions, respectively. The output from IP$^{-1}$ is the 64 bit cipher text block.

### 2.2 Triple Data Encryption Standard (TDES)

A DES algorithm is no longer considered to be a secure algorithm for many applications by the NIST (National Institute of Standard and Technology). A more secure algorithm based on DES is called Triple Data Encryption Algorithm (triple DES, 3DES, or TDEA) which is still supported by NIST. Fig. 4 shows the Triple Data Encryption Algorithm. This involves applying DES, then DES$^{-1}$, followed by a final DES to the plain text using three different options [7]. This results in a cipher text that is much harder to break. TDEA uses the same set of operations as DES.

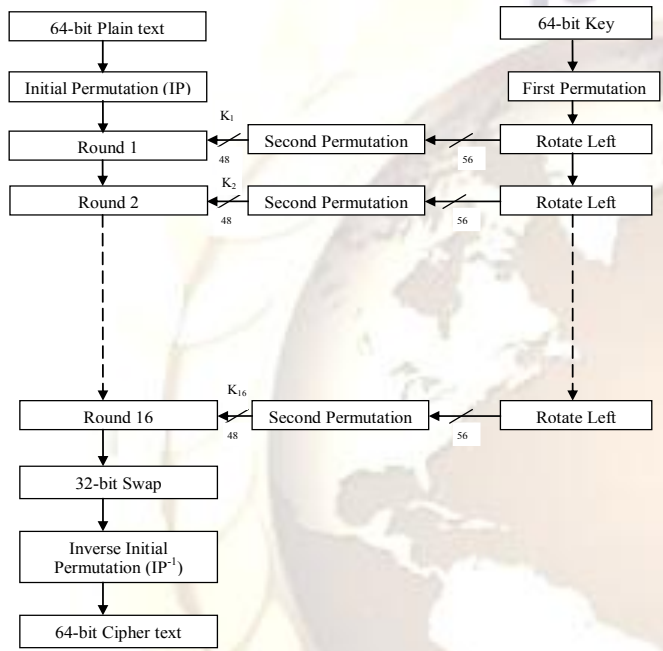

**Fig.2.** DES Algorithm

At any given step in the process, the new L block value is merely taken from the prior R block value. The new R block is calculated by taking the bit-by-bit exclusive-OR (XOR) of the prior L block with the results of applying the DES cipher function f, to the prior L block and $K_n$. ($K_n$ is a 48 bit value derived from the 64 bits DES key. Each round uses a different 48 bits according to the standards key schedule algorithm).

The cipher function, f, combines the 32-bit R block value and the 48-bit sub key in the following way. First, the 32-bits in the R-block and expanded to 48 bits by an expansion function (E); the extra 16 bits are found by repeating the bits in 16 predefined positions. The 48-bit expanded R block is then XORed with the 48-bit value that is then divided into eight 6-bit blocks. There are fed as input into 8 sections (S) boxes, denoted S1,…, S8. Each 6bit input yields a 4-bit output using a lookup table (LUT) based on the 64 possible inputs; this results in a 32-bit output from the S-box. The 32-bits are then arranged by a permutation function (P), producing the results from the cipher function.

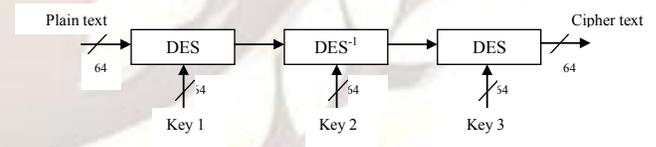

**Fig. 4.** TDES Block Representation

### 2.3 Advanced Encryption Standard (AES)

There are numerous encryption algorithms that are now commonly used in computation, but U.S government has adopted the Advanced Encryption Standard (AES) to be used by Federal departments, and agencies for protecting sensitive information. The AES algorithm is a symmetric cipher and used a single secret key for both the encryption and decryption. In addition, the AES algorithm is a block cipher as it operates on fixed-length groups of bits (blocks), whereas in stream ciphers, the plaintext bits are encrypted one at a time, and the set of transformation applied to successive bits may very during the encryption process. The AES algorithm operates on block length [$N_b$] of 128-bits, by using cipher keys with key length [$N_k$] of 128, 192 or 256 bits or the encryption process.

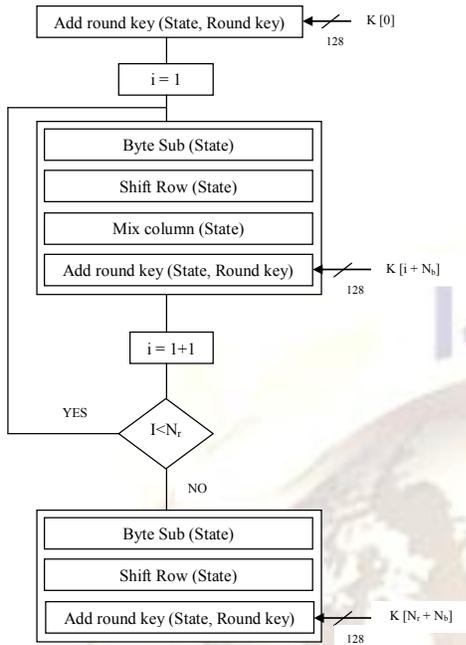

**Fig.5.** AES Block Diagram for Key Length of 128bits and the number of Iterations required are 10(Nr = 10)

The encryption and decryption process of AES block consists of number of different transformation applied consecutively over the data block bits, considered as a 4x4 array of 8 bit bytes (also called "state" in the algorithm). The state undergoes four different transformations in each round having fixed number of iterations. These transformations are *"Sub Byte"*, *"Shift Row"*, *"Mix Column"*, and *"Add Round Key"* transformations. *"Sub Byte"* can be implemented by non–linear substitution of bytes that operates independently on each byte of the state using a substitution LUT (S-box). In this S-box; each byte in the state matrix is an element of a Galois Field GF ($2^8$), with irreducible polynomial m(x) = $x^8 + x^4 + x^3 + x+1$. In simple terms, GF ($2^n$) is a set of $2^n$ elements each represented by an n-bit string of 0's and 1's and affine transformation is applied (over GF (2)). The *"Shift Row"* can be implemented using a cyclically shift the rows of the state over different offsets. *"Mix Column"* are considered as most complicated operation in the algorithm and need GF ($2^8$) fields and multiply by modulo $x^4+1$ with a fixed polynomial a(x)={03}$x^3$ + {01}$x^2$ + {02}x. *"Add Round Key"* is added to the state by a logical XOR operation. Each round key consists of $N_b$ words from the key expansion. These $N_b$ words are added into the state columns. Each round key is a 4-word (128bit) array generated as a product of previous round key, and a sense of substitution LUT for each 32-bit word of the key. The key expansion generated a total of $N_b(N_r+1)$.

## 3. Design and Implementation Methodology

Current applications demand high speed processor for large amount of data transmission in real time. As compared to software alternatives, hardware implementation provides highly secure algorithms and fast solutions approaches for high performance applications. Software approaches could be a good choice but it has some limitations like low performance and speed. Main advantages of software are low cost and short time to market. But they are unacceptable in terms of high speed and performance specification. So that, Hardware alternatives could be selected for implementing MIPS crypto processor architecture.

**Table 1.** Hardware v/s software alternatives for crypto processor

| Parameters | Software | Hardware | |
|---|---|---|---|
| | | FPGA | ASIC |
| Performance | Low | Medium-High | Very High |
| Power consumption | Depends | Very high | Low |
| Logic integration | Low | Low | High |
| Tool cost | Low | Low | Low |
| Test development complexity | Very low | Very low | High |
| Density | High | Very low | High |
| Design efforts | Low-medium | Low-medium | High |
| Time consumed | Short | Short | High |
| Size | Small-medium | Small | Large |
| Memory | Fine | Fine | Fine |
| Flexibility | High | High | - |
| Time to market | Short | Short | High |
| Run time configuration | - | high | - |

Hardware implementation supports both Field Programmable Gate Arrays (FPGAs) and Application Specific Integrated Circuits (ASIC) at high data rates. Such design has high performance but more time consuming and expensive as compared to software alternatives. The detailed comparison of hardware v/s software solutions for implementing the MIPS crypto processor architecture is shown in Table 1. Based on the comparison, hardware solution is a better choice in most of the cases because they have high performance. The main advantage of FPGA in hardware alternative, FPGA are low density and low area consumption. Logic integration, size and density are the major drawbacks in ASIC but have higher performance than FPGA.

### 3.1 Implementation of cryptographic engine

The global architecture of encrypted and decrypted MIPS pipeline processor is modified in a way that it executes encrypted instruction. Fig. 6 shows the block diagram of encrypted MIPS processor. To modify MIPS processor for encryption, we insert the cryptography module such as Data

Encryption Standard (DES), Triple Data Encryption Standard (T-DES), Advanced Encryption Standard (AES) etc. to the pipeline stage. Only single cryptographic module is used in same hardware implementation. The instruction fetch unit of encrypted MIPS contains Program Counter (PC), Instruction Memory, Decryption core and MUX. The Instruction memory reads address from the PC and stores instruction value at the particular address that is pointed by the PC. Instruction Memory sends encrypted instruction to MUX and decryption core. The decryption core gives decrypted instructions which are further sent to the MUX. The output of MUX is fed to the IF register. The MUX control signal comes from control unit. The instruction decode unit contains Register file and Key register. Key register stores the key data of encryption/decryption core. Key address and Key data comes from write back stage. The key data to be stored into the register file and remains same for all program instruction execution. The control unit provides various control signals to other stages. This acts as select line for two multiplexers. When the control unit detects a store/branch/jump it asserts the control signal high and keep it asserted till a load instruction is detected. During that period, the write back stage gets the forwarded data and the memory stage gets a constant zero value thus preventing only further transitions. When the control signal is de-asserted, then the data pass through the standard pipeline structure. The execute unit executes the register file output data and performs the particular operation determined by the ALU. The ALU output data is sent to EXE register which temporarily store address value. The Memory Access Unit contains Encryption core, Decryption core, Data Memory, MUX and DEMUX. The second register data from register file is fed to the encryption core and the MUX. Here the crypt signal enable/disable encryption operation when occurs. The read/write signal of data memory describes whether reading/writing operation is done. Output of data memory pass through DEMUX whose one output goes to decryption core and other to MEM register. Here the unencrypted memory data and decrypted data are temporarily stored to the MEM register. The MEM output is fed to the write back data MUX and according to the control signal, the output of MUX goes to register file.

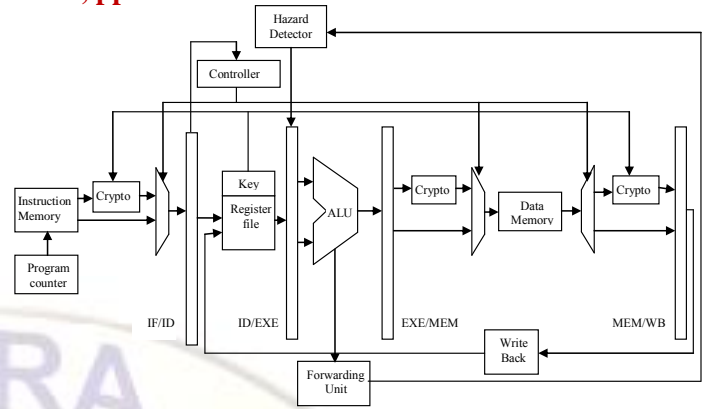

**Fig. 6.** Detailed MIPS crypto processor architecture

### 3.2 Microinstruction set

The MIPS instruction set is straightforward like any other RISC designs. MIPS are a load/store architecture, which means that only load and store instructions access memory. Other instructions can only operate on values in the registers [8]. Generally, the MIPS instructions can be broken into three classes: the memory-reference instructions, the arithmetic-logical instructions, and the branch instructions. Also, there are three different instructions formats (as shown in Fig.7) in MIPS architecture: R-Type instructions, I-Type instructions, and J-Type instructions. A subset of the instruction has been implemented in our design, the list of which is given in Table 2.

| Instruction Type | Instruction |
|---|---|
| R-Type | AND, OR, NOR, ADD, SUB, SLT |
| I-Type | ADDI, SUBI, NORI, ANDI, SLTI, SLL, SRL |
| | LW, SW, LKLW, LKUW |
| | BEQ, BNE |
| J-Type | J, JR, JAL, CRYPT |

**Fig. 7.** Implemented MIPS Instruction Types

**Table 2.** MIPS Instruction Format

R-Type: | Op | RS | RT | RD | Shamt | Funct |

*Arithmetic instruction format*

I-Type: | Op | RS | RT | Address/immediate |

*Transfer, branch, immediate*

J-Type: | Op | Target address |

*Jump instruction*

| Field | Description |
|---|---|
| Op[31-26] | 6-bit operation code |
| RS[25-21] | 5-bit source register |
| RT[20-16] | 5-bit target register |
| Immediate[15-0] | 16-bit immediate address |
| Target[25-0] | 26-bit jump target |
| RD[15-11] | 5-bit destination register |
| Shamt[10-6] | 5-bit shift amount |
| Funct[5-0] | 6-bit function field |

**Table 3.** ISA overview of Implemented MIPS Crypto Processor

| Instruction | Operation | Syntax | Opcode | Clock cycle |
|---|---|---|---|---|
| ADD | Arithmetic Addition operation | *Add $rd, $rs, $rt* | 0(0x000000) | 4 |
| SUB | Arithmetic Subtraction operation | *Sub $rd, $rs, $rt* | 0(0x000000) | 4 |
| AND | Logical AND operation | *AND $rd, $rs, $rt* | 0(0x000000) | 4 |
| OR | Logical OR operation | *OR $rd, $rs, $rt* | 0(0x000000) | 4 |
| NOR | Logical NOR operation | *NOR $rd, $rs, $rt* | 0(0x000000) | 4 |
| SLT | Set less than manipulation operation | *SLT $rd, $rs, $rt* | 0(0x000000) | 4 |
| ADDI | Immediate arithmetic addition operation | *ADDI $rd, $rs, constant* | 8(0x001000) | 4 |
| SUBI | Immediate Arithmetic Subtraction operation | *SUBI $rd, $rs, constant* | 8(0x001000) | 4 |
| SLTI | Immediate Set less than manipulation operation | *SLTI $rd, $rs, constant* | 8(0x001000) | 4 |
| ORI | Immediate Logical OR operation | *ORI $rd, $rs, constant* | 8(0x001000) | 4 |
| ANDI | Immediate Logical AND operation | *ANDI $rd, $rs, constant* | 8(0x001000) | 4 |
| NORI | Immediate Logical NOR operation | *NORI $rd, $rs, constant* | 8(0x001000) | 4 |
| SLL | Shift left logic operation | *SLL $rd, $rs, shamt* | 0(0x000000) | 4 |
| SRL | Shift right logic operation | *SRL $rd, $rs, shamt* | 0(0x000000) | 4 |
| BEQ | Branch equal operation | *Beq $rd, $rs, label* | 4(0x000100) | 3 |
| BNE | Branch not equal operation | *Bne $rd, $rs, label* | 4(0x000100) | 3 |
| JR | Conditionally jump to register | *Jr $rd* | 2(0x000010) | 3 |
| JAL | Unconditionally jump to program | *Jal $rd* | 2(0x000010) | 3 |
| J | Conditionally jump to program | *J $rd* | 2(0x000010) | 3 |
| CRYPT | Encryption/decryption enable | *Crypt $rd* | 65(0x111111) | 3 |
| LW | Load data word to CPU | *Lw $rd, offset($rs)* | 35(0x100011) | 5 |
| SW | Store data to memory | *sw $rd, offset($rs)* | 43(0x101011) | 4 |
| LKUW | Load key upper word to target register | *LKUW $rd, offset($rs)* | 64(0x111110) | 5 |
| LKLW | Load key load word to target register | *LKLW $rd, offset($rs)* | 60(0x111100) | 5 |

The MIPS instruction Format is used to minimize the number of bits in each instruction, note that the 6-bit operation code field in the instruction format is used to have the word length of the memory as 32-bit and used standard memory blocks for the program memory. Only 32-bit instruction set is required for the current implementation as shown in Table 3. There are three more new instructions that support encrypted and decrypted operation. These instructions are load key upper word (LKUW), load key lower word (LKLW) and encryption mode (CRYPT).

These instructions randomly use opcodes in the hardware implementation. LKLW and LKUW come under I-type instruction and variant of load word (LW). These two instructions do not need to specify a destination address in the assembly code. CRYPT instruction comes under J-type instruction and instead of address, only single argument i.e., Boolean value is assigned. This indicates enable/disable encryption and decryption process. Any non zero value enables the encryption/decryption process and zero value disables the encryption process.

### 3.3 Initialization
The operational mode of the MIPS crypto processor is controlled by a RESET signal. When the RESET signal is at logic "0", the crypto processor is in the reset mode and the processing unit writes the memory and register contents using the 32-bit bidirectional data bus, 10-bit address bus, and four control signals. When the reset signal is at logic "1", the crypto processor is in the running mode and acts as an FPGA, implementing one of the three algorithms based on the preloaded contents of the memory blocks. The keys are kept in the key registers of the register file of crypto processor that are available to other stages of processor.

### 4. Implementation Results
The complete pipeline processor stages are modeled in VHDL. The syntax of the RTL design is checked using Xilinx tool. For functional verification of the design the MIPS processor is modeled in Hardware Descriptive Language. The design is verified both at the block level and top level. The complete design along with all timing constraints, area utilization and optimization options are described using Synthesis Report. The design has been synthesized targeting 40nm triple oxide process technology using Xilinx FPGA Virtex-6 (xc6vlx240t-3ff1156) device. The Virtex family is the latest and fastest FPGA which aims to provide up to 15% lower dynamic and static power and 15% improved performance than the previous generation [18]. It is obvious that there is a trade-off between maximum clock frequency and area utilization (number of slices LUT's) because the basic programmable part of FPGA is the slice that contains four LUTs (look up table) and eight Flip flops. Some of the slice can use their LUT's as distributed RAM.

### 4.1 Power Reduction Technique
One of the key concerns in any microprocessor based system is power consumption. Power dissipation is either static or dynamic. Static power dissipation is caused due to the leakage and short circuit current while dynamic power dissipation is due to switching activity of the various transistors in the circuit. Dynamic power forms the major chunk of power dissipation in CMOS circuits and have require a lot of attention. In our design power reduction is achieved through bypassing pipeline stages that cause unnecessary switching activity. One of the influencing factors of dynamic power dissipation is switching activity and dynamic power is given by the equation,

$$P = 0.5 \, C \, (V_{dd})^2 \, E \, (sw) \, F_{clk}$$

There by decreasing switching activity (E (sw)) results in reduced dynamic power consumption. The pipeline stages for different type of instructions mentioned above are shown in Fig. 8. It can be seen that the data memory stage of the pipeline is not used by any of the arithmetic instructions. Transition during this unused state causes extra power dissipation. To avoid this wastage, the pipeline is reconfigured to bypass this stage for these set of instructions. Hence data obtained from execution stage is forwarded directly to the write back stage. During this time, the EXE/MEM pipeline registers are maintained at zero value thus ensuring that no transition take place and power dissipation is reduced.

Arithmetic instruction has a 'NOP' stage in the MEM stage while there is a 'NOP' during the write back stage for the Store/Branch/Jump instructions. Hence the write back stage of the arithmetic could be moved to the MEM stage without causing any resource conflicts. A load instruction used all five stages of the pipeline and hence a resource conflict will arises. So, data has to pass through the regular pipeline structure till it encounters a store/branch/jump instruction after which the reconfigured pipeline can be again brought in.

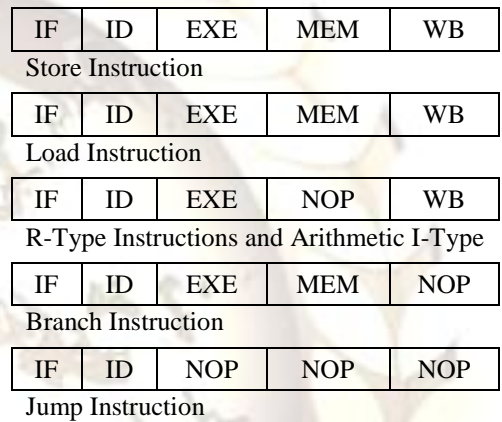

| IF | ID | EXE | MEM | WB |

Store Instruction

| IF | ID | EXE | MEM | WB |

Load Instruction

| IF | ID | EXE | NOP | WB |

R-Type Instructions and Arithmetic I-Type

| IF | ID | EXE | MEM | NOP |

Branch Instruction

| IF | ID | NOP | NOP | NOP |

Jump Instruction

**Fig. 8.** MIPS Instruction Format

### 4.2 Performance, Area and Power Requirement
The performance of MIPS crypto processor based on three different crypto modules such as DES, TDES, and AES algorithms. For DES and TDES, 16 clock cycles are used for DES/TDES crypto specific block to execute data, 20 clock cycles are needed to execute the R-type instruction, 21 clock cycle are needed for I-type instruction and 19 clock cycle for J-type instruction data.

The power consumption is estimated by the Xilinx XPOWER Analyser tool, using the post layout netlist of the crypto processor along with the node activity data for each algorithm. The power consumption can be further reduced by running the processor at lower voltages than the normal voltage of 1.5v (as long as the speed and throughput requirements are satisfied). Power analysis was done for the portion between the EXE/MEM and MEM/WB stage. This is performed for both the encryption and decryption process. Clock gating technique is used to minimize energy reduction during pipeline stall stages. This technique identifies low processing requirement periods and reduces operating voltage with clock frequency (voltage-frequency scaling), resulting in reduced average operating power consumption. This may or may not occur frequently depending upon compiler efficiency. The power analysis result is carried out on the same clock frequency. in our design, a symbol is processed every clock cycle, the throughput is calculated on the basis of number of instruction execution per second. The formula for calculating throughput is:

Throughput = f * symbol width/total clock frequency

Where f is the operation frequency and symbol width is one of our parameterized values. Table 3 and Table 4 show the performance throughput, area and the estimated power consumption of DES and TDES MIPS crypto processor. Maximum throughput of MIPS DES based crypto processor is of 664Mbits/s at 4.58ns and for TDES based crypto processor is 636Mbits/s at 4.78ns.

**Table 3.** Throughput estimation of MIPS crypto processor (DES Based)

| Features | Processor |
|---|---|
| Crypto processor | DES |
| Data length | 64-bits |
| Speed | 218MHz (clock rate) |
| Throughput | 664 Mbits/s (Data Bandwidth) |
| Area | 66072 Slice LUT's(look up tables) |
| Latency | 21 clock cycles(both for encryption and decryption) |
| Power consumption | 1.746W(quiescent-1.303 and dynamic-0.444) |

**Table 4.** Throughput estimates for the MIPS crypto processor (TDES Based)

| Features | Processor |
|---|---|
| Crypto processor | TDES |
| Data length | 64-bits |
| Speed | 209MHz(clock rate) |
| Throughput | 636 Mbits/s (Data Bandwidth) |
| Area | 64673 Slice LUT's(look up tables) |
| Latency | 21 clock cycles(both for encryption and decryption) |
| Power consumption | 1.981W(quiescent-1.131 and dynamic-0.851) |

In case of AES crypto processor, 43 clock cycles are used for crypto specific block to execute data, 47 clock cycles are needed to execute the R-type instruction, 48 clock cycles are needed for I-type instruction and 46 clock cycles for J-type instruction data. Table 5 shows the performance throughput; area and the estimated power consumption of AES based MIPS crypto processor. Maximum throughput of AES based MIPS Crypto processor is 560Mbits/s at 4.76ns. Moreover, it is possible to trade performance with area and power in the implementation. For example, higher performance can be obtained by running processor at higher frequency up to 300MHz for the current design (increasing power consumption) and/or using pipeline (increasing area) for more performance demanding applications.

**Table 5.** Throughput estimates for the MIPS crypto processor (AES Based)

| Features | Processor |
|---|---|
| Crypto processor | AES |
| Data length | 128-bits |
| Speed | 210MHz (clock rate) |
| Throughput | 560Mbits/s (Data Bandwidth) |
| Area | 109738 Slice LUT's(look up tables) |
| Latency | 48 clock cycles(for encryption) |
| Power consumption | 1.313W(quiescent-1.008 and dynamic-0.396) |

## 5. Discussion
In this section, we address a few extra features that can be added to the design in the future to enhance its capabilities. In the current design, we have mainly concerned about the feasibility, area requirement and the performance of the MIPS crypto processor, which is important for high speed applications and less concerned about the power consumption. This is a constraint specific to security application only. However, we have used several implementation techniques, such as using keeper cells on the bus lines and clock gating for the registers in register file and memory unit to keep the power consumption low. In a more area limited application, the memory size can be reduced to 128x32 bits. This can execute all those instructions which are required for all cryptography algorithms. To reduce power consumption, the following techniques are available to the designers: (1) Reducing the bus wire lengths by using a hierarchical implementation, instead of a flat implementation and (2) Using a selective clock signal for the registers and memory to clock the active portion of the registers for cases where operands occupy less than the full register width. (3) When a bus is not going to be used in a Datapath, it is held in a quiescent state by stopping the propagation of switching activity through the module driving the bus.

## 6. Conclusion
In this paper, we have presented an efficient hardware architecture design of 32-bit encrypted and decrypted MIPS processor that executes encrypted/decrypted instructions. Initially it read encrypted data from instruction memory and decrypts the same data and sent it to the next pipeline stages. The processor uses the symmetric block viz., DES, TDES and AES plain/cipher that can process data length of 64bits and 128bits respectively. The crypto block in the MIPS processor performs data encryption and decryption. The design has been modeled in VHDL and functional verification policies are adopted for it. Optimization and synthesis of design is carried out at latest and fastest FPGA Viretx-6 device that improves performance. Each program instructions are tested with some of vectors provided by MIPS. We conclude that the performance of MIPS crypto processor using DES and AES is High 664Mbits/s and 560Mbits/s respectively. The high performance and high flexibility of crypto processor design makes it applicable to various security applications.

## Biography

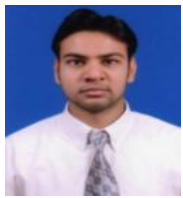

Kirat Pal Singh is Pursuing Master of Technology in VLSI Design from Centre for Development of Advanced Computing, Mohali since 2010 and received the Bachelor Degree in Electronics and Communication Engineering from Punjabi University, Patiala, in 2010. His current research interests include Digital system design, Embedded Systems and System-on-Chip (SOC) Technology, VLSI and Microelectronics System Design, Hardware Crypto System Design for Computer Security. He has published an extensive paper on these subjects in international conference proceedings and Journals.

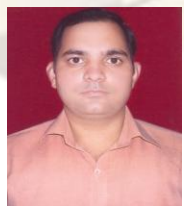

Dr. Dilip has done his Ph.D degree in the area of Wireless Sensor Networks. He obtained his Master of Engineering (M.E.) from the Department of Electronics & Electrical Communication, PEC University of Technology, Chandigarh in 2003 and Bachelor of Engineering (B.E.) from the Department of Electronics & Telecommunication, Army Institute of Technology, University of Pune, Pune in 2000. He is a Senior Design Engineer at Centre for Development of Advanced Computing (C-DAC), Mohali. His research interests include Embedded Systems, Power

Electronics, Electronics Product Design, Wireless Sensor Networks and Analog & Digital Electronics etc. He has published more than 50 high quality research papers on these subjects in International journals and conferences viz. Elsevier, IEEE, ACM, Inderscience etc.

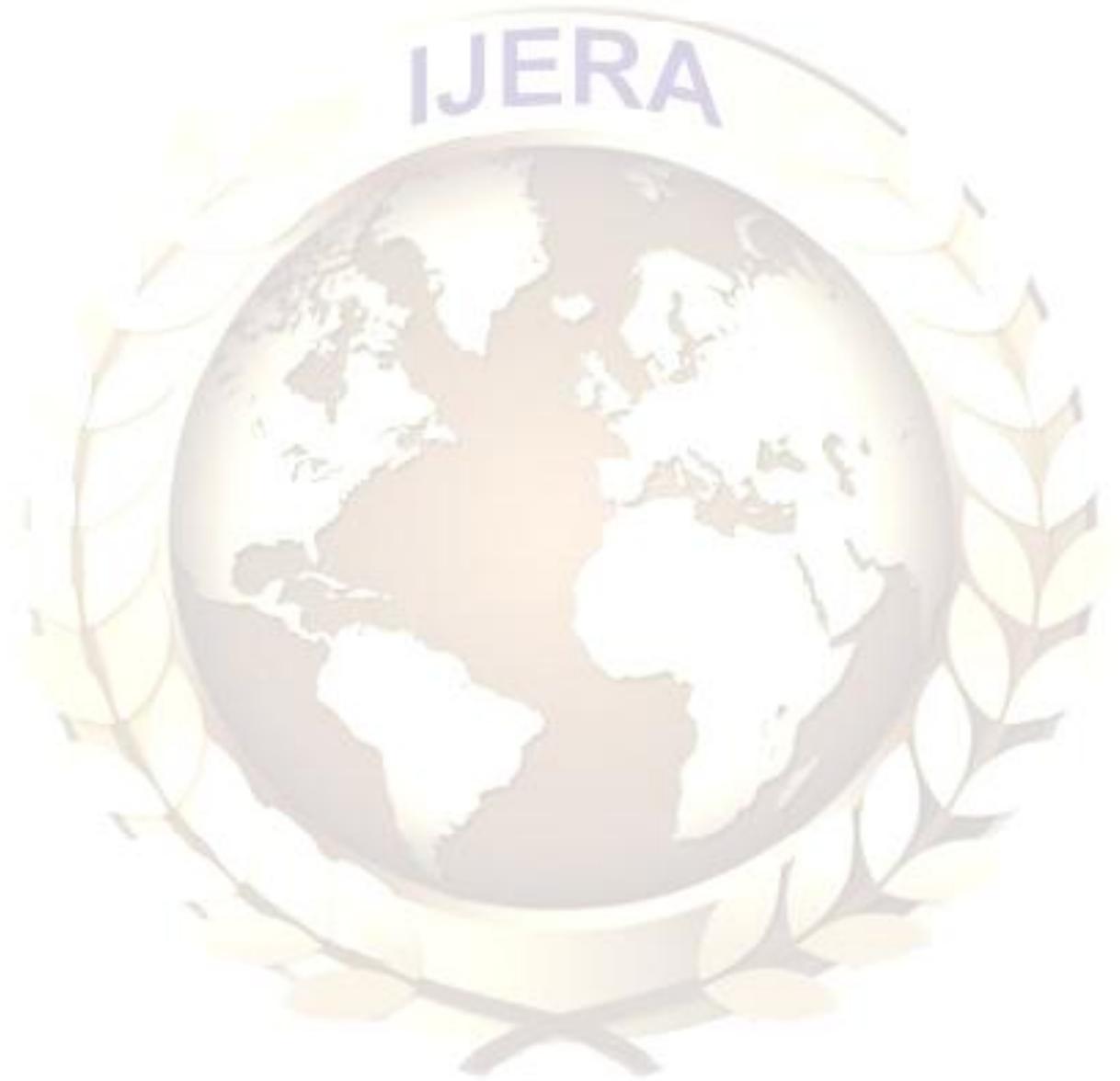